\documentclass[a4paper,11pt]{article}
\pdfoutput=1 

\usepackage{booktabs}
\usepackage{todonotes}
\usepackage{jheppub} 

\usepackage[T1]{fontenc} 

\graphicspath{{figures/}{build/figures/}}

\title{Precision Cell Resampling with a Relative and Resonant Aware Metric}

\preprint{IPPP/26/38}

\author[a]{Jeppe R.~Andersen,}
\author[b]{Ella Cole,}
\author[c,d]{and Andreas Maier}

\affiliation[a]{Institute for Particle Physics Phenomenology, University
  of Durham, South Road, Durham DH1 3LE, UK}
\affiliation[b]{DAMTP, University of Cambridge, Wilberforce Road, Cambridge, CB3 0WA, United Kingdom}
\affiliation[c]{%
The Henryk Niewodniczański Institute of Nuclear Physics,
ul. Radzikowskiego 152, 31-342 Krakow, Poland
}
\affiliation[d]{Institut de F{\'i}sica d'Altes Energies (IFAE), The Barcelona Institute of Science and Technology, Campus UAB, 08193 Bellaterra (Barcelona), Spain}

\emailAdd{jeppe.andersen@durham.ac.uk}
\emailAdd{egc41@cam.ac.uk}
\emailAdd{andreas.maier@ifj.edu.pl}

\abstract{%
  We present a metric on the space of scattering events based on
  relative transverse momenta and with explicit sensitivity to
  intermediate resonances. With this new metric, negative weights in an
  event sample can be reduced substantially through cell resampling,
  while preserving the predicted properties of the resonance with high
  accuracy. We demonstrate the efficiency on a NLO event sample for the
  production of a leptonically decaying W boson together with two jets.
}

\begin{document}
\maketitle
\flushbottom

\section{Introduction}
\label{sec:intro}

The focus of the physics programme at the LHC is increasingly turning towards precision measurements.
The interpretation of data requires high-statistics simulated event samples based on increasingly sophisticated theory predictions.
The required increases in
systematic precision and in the sample size both inflate the
computational costs. Already today, event simulation makes up a
substantial fraction of the ATLAS and CMS computing
budgets. Significant improvements in the efficiency of event generation will
be crucial to ensure the uncertainty on the predictions from theory can match the uncertainty in the experimental measurements. This will be true especially after the
High-Luminosity (HL-LHC) upgrade.

Predictions based on higher-order perturbation theory not only require
more computing time for the simulation of each individual event, but
additionally tend to inflate the number of events needed to reach a
given statistical accuracy. The primary cause is counter-events with
negative Monte Carlo weights. Reducing the fraction of such
negative-weight events can accelerate the statistical convergence by
orders of magnitude. Hence, there have been considerable efforts in
suppressing the impact of negative-weight events.

For parton-shower matched predictions negative weights can be
suppressed or even completely eliminated by developing a suitable
formalism for the combination with fixed-order predictions, at least
up to
NLO~\cite{Frixione:2007vw,Jadach:2015mza,Danziger:2021xvr,Nason:2021xke,Frederix:2023hom,Shyamsundar:2025nzn,Shyamsundar:2025mfw,vanBeekveld:2025lpz,Farkh:2026mtw}. A
complementary approach is to identify groups of similar events with
mixed positive and negative weights and to redistribute these weights
in a way that preserves observables and reduces negative
weights~\cite{Andersen:2020sjs,Nachman:2020fff,Andersen:2021mvw,Andersen:2023cku,Nachman:2025lid,Palmer:2025jmb}. The
\emph{cell resampling} method proposed in~\cite{Andersen:2021mvw}
applies to both fixed-order and
parton-showered~\cite{Andersen:2024mqh} event generation and can be
used either a posteriori on the final event sample or to optimise the
generation itself~\cite{Ulrich:2025fij}. It is essential that the
notion of similarity, i.e.~the distance assigned to pairs of events,
matches the sensitivity of real-world experimental analyses. The reweighing should
strive to preserve regions of high statistical and systematic accuracy
as best as possible, while larger distortions in less sensitive
phase-space regions may be acceptable. A particular challenge is posed
by intermediate resonances, whose properties do not directly enter the
metric originally suggested in~\cite{Andersen:2021mvw}.

In the following, we show that with a suitably chosen distance
function cell resampling preserves distributions of an intermediate
resonance at the sub-per cent level, while substantially reducing the
contribution from negative weights. In section~\ref{sec:metric}, we
review the metric introduced in~\cite{Andersen:2021mvw} and propose
improvements to better reflect the sensitivities of experimental
analyses. We then assess the performance for the production of a W
boson with two jets at NLO in section~\ref{sec:Wjj}. We conclude in
section~\ref{sec:conclusions}.

\section{A Relative and Resonant Metric}
\label{sec:metric}

Cell resampling at its core is the organisation of similar events into
sets (or cells) where weights can be redistributed. The method is
described in detail in~\cite{Andersen:2021mvw,Andersen:2023cku}. In
short, one selects an event with negative weight as cell seed,
constructs a cell by adding a number of nearest neighbours according
to some distance function, and averages weights between the
events inside the cell. The procedure is repeated for each
negative-weight event in the original sample. Clearly, the distance
function (or metric) plays a central role, determining the events to
be added to a cell. In the following, we first review the original
``absolute'' metric introduced in \cite{Andersen:2021mvw} in
section~\ref{sec:metric} and then suggest two modifications in
section~\ref{sec:metric_rel}.

\subsection{The Absolute Metric }
\label{sec:metric_abs}

To determine the similarity between two events $e$ and $e'$ we first
define the observable objects, e.g.\ jets, (dressed) leptons, and
isolated photons. In~\cite{Andersen:2021mvw}, the distance between these
events is then defined as the sum over the distances among objects of
the same type $t$, viz.
\begin{equation}
  \label{eq:dist_events_l1}
  d(e, e') = \sum_{t=1}^T d(s_t, s_t').
\end{equation}
$s_t$ and $s_t'$ represent the set of objects of type $t$ in $e$ and
$e'$, respectively. We ensure that their cardinalities are the same by
adding auxiliary objects with vanishing momentum. To determine the
distance between the sets of objects of type $t$ we consider pairings
$p_i$, $p_j'$ of these objects between $e$ and $e'$ and sum over the pairwise
distances. We select the pairing that is optimal in the sense of
minimising this sum. That is
\begin{equation}
  \label{eq:dist_set_l1}
  d(s_t,s_t') = \min_{\sigma \in S_P} \sum_{i=1}^P d_{\text{abs}}(p_i, p_{\sigma(i)}'),
\end{equation}
where $S_P$ is the group of all permutations of the $P$ objects of the
given type $t$. Finally, the distance between two objects of the same
type with momenta $p$ and $p'$ is based on the absolute distance
between the spatial momenta:
\begin{equation}
  \label{eq:dist_abs}
d_{\text{abs}}(p, p') = \sqrt{\sum_{i=1}^3 (p_i-p_i')^2 + \tau^2(p_\perp - p_\perp')^2}.
\end{equation}
The tunable parameter $\tau$ is introduced to account for a
greater sensitivity to the transverse momentum components $p_\perp, p_\perp'$.

\subsection{A Relative Metric with Intermediate Resonances}
\label{sec:metric_rel}

The absolute metric discussed in section~\ref{sec:metric_abs} works
best for preserving observables that are directly related to the
properties of the observed objects entering the distance function. For
observables based on composite objects, for instance intermediate
resonances reconstructed from their decay products, one typically
finds somewhat larger distortions. As an example, for the diphoton
production considered in~\cite{Andersen:2024mqh} the local deviation
from the original sample never rises above a few per mille and always
stays well within one standard deviation in the transverse momentum
spectrum of the hardest photon when limiting the cell radius to
$30\,$GeV. In contrast, the largest change in the diphoton spectrum
amounts to almost $5\%$. What is more, big effects are found at small
transverse momenta, where the statistics are highest and the
resolution of the experimental analysis is best. Indeed, in this
kinematic region we observe differences of up to seven standard
deviations.

Our first goal is to preserve the properties of resonances with the
same accuracy as those of the final-state objects. To this end, we
reconstruct the momentum of the resonance and, for the purpose of
computing distances, treat it like a new type of observable
object. Our next goal is then to better match experimental resolution
to ensure higher accuracy in regions where it is needed. Here, we
observe that the distance between two objects as defined in equation
\eqref{eq:dist_abs} is based on the \emph{absolute} spatial momenta,
whereas experimental resolution is better modelled \emph{relative} to
the energy of the particle. For example, in recent ATLAS analyses, the
energy scale uncertainty for a typical jet ranges from $\sim 40\%$ at
low transverse momenta to $5\%$ at high transverse
momenta~\cite{ATLAS:2020cli}. Similarly, relative uncertainties are
obtained for the photon energy. Furthermore, for many distributions
fewer events may be observed at large energy necessitating the use of
logarithmic binning.  For these reasons, we may wish to use a relative
metric inside the cell resampler, such that the cell size is typically
much smaller where an observable is precisely measured and large
elsewhere, ensuring that the cell size increases (and therefore allows
for a larger cancellation of negative weights) where that is
commensurable with the required accuracy of the description. This
implies larger cells in regions where events are less dense and
provide fewer opportunities for weight cancellation.

When constructing a cell resampling distance it is crucial to fulfil the
requirements of a metric, both for soundness~\cite{Andersen:2021mvw}
and to be able to use efficient nearest-neighbour search algorithms,
e.g.~the vantage-point tree search discussed
in~\cite{Andersen:2023cku}. To this end, we consider
\begin{align}
  \label{eq:dist_rel}
d_{\text{rel}}(p,p') = \alpha\, \left\lvert\log\frac{p_\perp}{p_\perp'}\right\rvert + \beta\, \Delta
  R(\vec p,\vec{p}\,')
\end{align}
with $\alpha, \beta > 0$.
$\Delta R = \sqrt{\Delta \phi^2 + \Delta y^2}$ is the distance in the
plane spanned by the azimuthal angle $\phi$ and the rapidity
$y$. $d_{\text{rel}}$ indeed defines a valid metric, see
appendix~\ref{sec:metricproperties} for details. For similar objects
with $\Delta R \approx 0, p_\perp \approx p_\perp'$ we measure
relative differences
$d_{\text{rel}}(p,p') \approx \alpha \lvert p_\perp - p'_\perp \rvert
/ p_\perp$. The exact behaviour for dissimilar objects is irrelevant
as long as we impose strict limits on the maximum cell size.

The parameters $\alpha$ and $\beta$ can be tuned to alter the relative
importance of the transverse momentum distance and $\Delta R$. In
principle, different values can be chosen for different object
types. For instance, better experimental resolution of hard photons
compared to jets would be an incentive to preserve predictions for
photon momentum distributions with a finer granularity. This could be
achieved by choosing larger values for $\alpha$ and $\beta$ for
photons. Leaving such considerations aside for the moment, we
universally use $\alpha = \beta = 1$ in the following, such that a
$10\%$ change in transverse momentum and $\Delta R = 0.1$ are counted
with approximately the same weight. Appropriate choices and
optimisations of these parameters are left for further studies.

Next, we lift the distance between individual objects to an event
distance. For the absolute distance, we added up distances linearly in
equations~\eqref{eq:dist_events_l1}, \eqref{eq:dist_set_l1}. For a
relative distance this combination procedure does not appear
well-motivated. Instead, we note that any $p$-norm
\begin{align}
  \label{eq:dist_events_lP}
  d_p(e, e') ={}& \left[\sum_{t=1}^T d(s_t, s_t')^p\right]^{1/p},\\
  \label{eq:dist_set_lP}
  d_p(s_t,s_t') ={}& \min_{\sigma \in S_P} \left[\sum_{i=1}^P d_{\text{rel}}(p_i, p_{\sigma(i)}')^p\right]^{1/p},
\end{align}
fulfils the requirements of a metric and choose the maximum norm $p
\to \infty$.

When comparing events with different multiplicities our conventional
approach was to add auxiliary particles with vanishing momenta. When
using the relative metric from equation~\eqref{eq:dist_rel} this will
lead to an infinite distance. Hence, the phase space is disconnected
and we instead apply cell resampling to each multiplicity separately.

\section{Cell Resampling for W Boson Plus Dijet Production at NLO}
\label{sec:Wjj}

To assess the real-world performance we apply cell resampling to NLO
event samples for the production of a positron and an electron
neutrino via a virtual W boson together with two jets. In
section~\ref{sec:metric_cmp}, we analyse the difference between the
absolute metric summarised in section~\ref{sec:metric_abs} and the
relative metric including intermediate resonances introduced in
section~\ref{sec:metric_rel}. We then show in section
\ref{sec:precision_cres} that the new metric can preserve predictions
at the sub-per cent level, well within uncertainties, while
significantly reducing the contribution from negative weights.

\subsection{Comparison of Absolute and Relative Metrics}
\label{sec:metric_cmp}

In the following, we compare the performance of the different metric
definitions introduced in \cite{Andersen:2021mvw} and the present
work. We stress that the settings are specifically chosen to highlight
the features of the various distance measures and we recommend that
\emph{real-world precision analyses should use smaller maximum cell
sizes} than we do in this section, c.f.\
section~\ref{sec:precision_cres}.

We generate $N = 2.4\times 10^8$ weighted events using Sherpa
2.2.16~\cite{Bothmann:2019yzt} with \mbox{OpenLoops}~\cite{Buccioni:2019sur}
with the parameters summarised in table~\ref{tab:sherpa_param}.

\begin{table}[htb]
  \centering
  \begin{tabular}{ll}
    \toprule
    Parameter & Value \\
    \midrule
    $\sqrt{s}$ & 13\,TeV \\
    $\mu_r,\mu_f$ & $M_{W\perp}$\\
    PDF & NNPDF 3.1~\cite{NNPDF:2017mvq,Buckley:2014ana}\\
    Jet definition & anti-$k_t$~\cite{Cacciari:2008gp}\\
              & R = 0.4\\
              & $p_\perp$ > 30\,GeV\\
              & $\lvert \eta \rvert$ < 4.4 \\
    \bottomrule
  \end{tabular}
  \caption{Parameters chosen for the generation of W boson plus dijet
    events at NLO. $M_{W\perp}$ is the transverse mass of the W boson
    as reconstructed from the positron and the neutrino.}
  \label{tab:sherpa_param}
\end{table}

Following the discussion in section~\ref{sec:metric_rel}, we split up
the event sample according to multiplicity and apply cell resampling
separately to each of the subsamples. The final samples are analysed
with Rivet 3.1.10~\cite{Bierlich:2019rhm} using the
\texttt{ATLAS\_2011\_I925932}~\cite{ATLAS:2011fpo} analysis.

Our baseline for cell resampling is set by adopting the absolute
metric introduced in~\cite{Andersen:2021mvw} and summarised in
section~\ref{sec:metric_abs}. In line with the findings of our
previous
studies~\cite{Andersen:2021mvw,Andersen:2023cku,Andersen:2024mqh} we
set $\tau = 10$ to enhance sensitivity to transverse momentum
differences. As observable objects entering the metric, we consider
jets defined according to table~\ref{tab:sherpa_param}, positrons with
a transverse momentum of at least $20\,$GeV, and a missing transverse
momentum if it exceeds $25\,$GeV. Usually, the goal of cell resampling
is to suppress negative weights while preserving predictions within
their uncertainties, which can typically be achieved by choosing a
maximum cell radius of the order of $10\,$GeV. However, our present
aim is to instead illustrate the limitations of the original metric,
c.f.~section~\ref{sec:metric_rel}. We therefore choose a comparatively
large maximum cell radius of $50\,$GeV, stressing that smaller maximum
cell sizes are strongly recommended for practical applications.

We compare the baseline metric to a ``resonant'' improvement, adding
the reconstructed W boson to the observed objects. For the W boson
reconstruction, we closely follow the \texttt{ATLAS\_2011\_I925932}
analysis. First, we define the missing spatial momentum
$\vec{p}_{\text{miss}}$ as minus the sum of the final state spatial
momenta excepting the neutrino. The missing energy is defined as
$E_{\text{miss}} = \lvert \vec{p}_{\text{miss}} \rvert$. The
reconstructed W boson candidate four-momentum is then taken as $p_W =
p_l + p_{\text{miss}}$, where $p_l$ is the positron momentum. The
reconstructed transverse mass\footnote{
transverse mass $m_{W\perp}$ is in general different from the actual
transverse mass $M_{W\perp}$ used to set the renormalisation and
factorisation scale.} $m_{W\perp}$ is obtained from
\begin{equation}
  \label{eq:mw_t} m_{W\perp}^2 = 2 p_{l\perp} p_{\text{miss}\perp} (1 -
\cos \phi),
\end{equation} where $\phi$ is the azimuthal angle between the
positron momentum and the missing momentum. We add a W boson with
momentum $p_W$ to the metric objects if and only if $m_W \equiv
\sqrt{p_W^2} < 1\,$TeV and $m_{W\perp} > 40\,$GeV.

As a final improvement, we define the ``resonant + relative''
metric, where we both reconstruct the W boson as before and use the
``relative'' metric introduced in section~\ref{sec:metric_rel} instead
of the ``absolute'' one. Here, we tune the upper limit for cell radii
to $d(e, e') = 0.2$, matching the negative weight reduction
obtained with the absolute metric including the reconstructed W boson.

\begin{figure}[htb]
  \centering
  \includegraphics[width=0.45\linewidth]{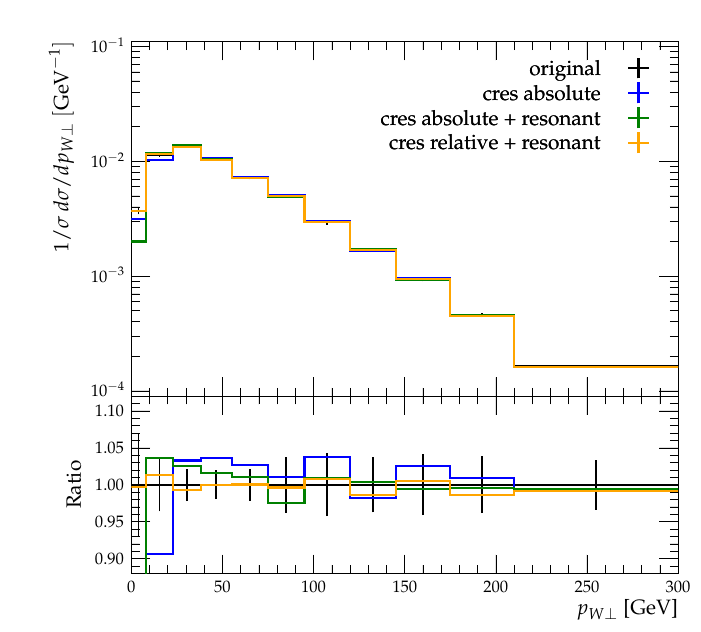}
  \includegraphics[width=0.45\linewidth]{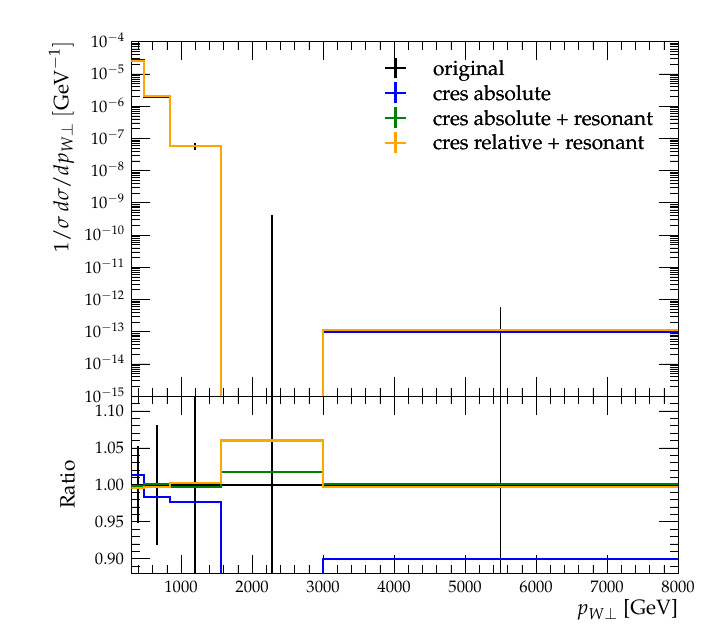}
  \caption{Predictions for the transverse momentum distributions of
the W boson before and after cell resampling. The left panel shows the
distribution binned according to the $\texttt{ATLAS\_2011\_I925932}$
Rivet analysis. The right panel showcases the behaviour at large
transverse momenta above 300\,GeV. The original distribution (black)
is compared to cell resampling using the conventional absolute metric
(blue), a ``resonant'' metric including a reconstructed W boson
(green), and the ``resonant + relative'' metric from
section~\ref{sec:metric_rel} (orange).  Artifically large maximum cell
sizes were chosen to highlight the differences between the considered
metrics.}
  \label{fig:ptW}
\end{figure}

Figure~\ref{fig:ptW} shows the distribution of the W boson transverse
momentum before and after cell resampling. With the
artificially chosen large maximum cell size the absolute metric --- with
or without resonant improvement --- leads to significant changes outside
the statistical uncertainties for small transverse momenta,
$p_{W\perp} < 25\,$GeV. It is of course expected that differences can
occur on scales smaller than the cell size. We further note that
directly including the W boson in the metric leads to a substantial
improvement in most bins, typically reducing the relative difference
to the original distribution by a factor of two or more.

Switching to the relative metric, we find much smaller
changes of about one per cent at small transverse momenta, well within
the statistical uncertainty. Conversely, the relative metric allows
larger cell sizes in the tail of the transverse momentum distribution,
which are expected to lead to a better suppression of negative
weights. Consequentially, we find the changes in this region to be
somewhat bigger, but still well within the statistical uncertainty.

\begin{figure}[htb]
  \centering
  \includegraphics[width=0.45\linewidth]{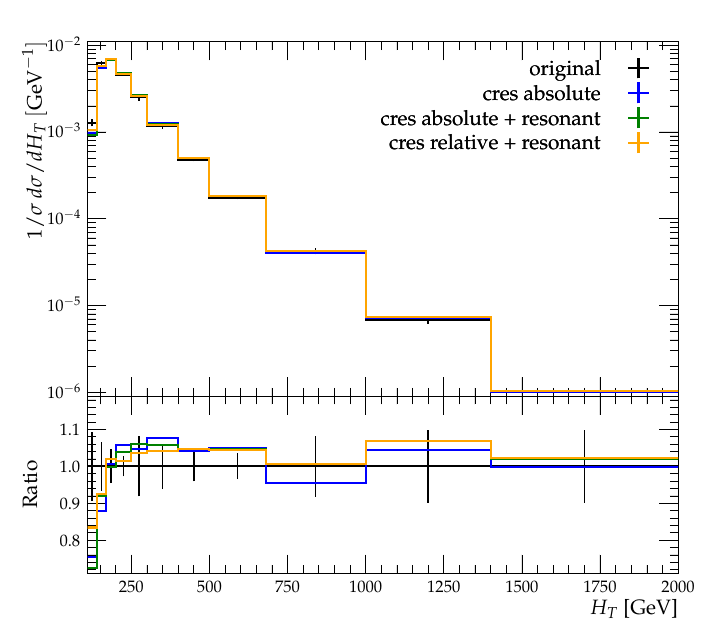}
  \caption{Predicted $H_T$ distribution before and after cell
    resampling. The original distribution (black) is compared to cell
    resampling using the conventional absolute metric (blue), a
    ``resonant'' metric including a reconstructed W boson (green), and the
    ``resonant + relative'' metric from section~\ref{sec:metric_rel}
    (orange). Artifically large maximum cell sizes were chosen to
    highlight the differences between the considered metrics.}
  \label{fig:HT}
\end{figure}

The superior behaviour of the relative metric for small transverse
momenta is further confirmed in distributions accounting for the
final-state partons. In figure~\ref{fig:HT} we show the distribution
in $H_T$, defined as the sum of the scalar transverse momenta of the
positron, the antineutrino, and all jets separated by $\Delta R > 0.5$
from the positron. This observable exhibits a sharp peak and is
therefore very sensitive to the cell sizes, since there are large
variations in the event weights over small distances in the metric. We
stress again that in this work we choose comparatively large maximum
cell sizes for illustration purposes and generally recommend smaller
limits for precision analyses. For $H_T > 170\,$GeV we find that the
absolute and relative metrics perform similar, inducing small changes
comparable to the statistical uncertainty in the original sample. In
contrast, for smaller values of $H_T$ the relative metric performs
much better, remaining visibly closer to the original distribution.

While negative weights are stochastically distributed in phase space,
there are systematic density fluctuations. Hence, cell resampling
induces both statistical and systematic changes in predictions.
To quantify these changes we consider the $\chi^2$ per degree of freedom
($\chi^2$/d.o.f.) difference from the original predictions. As we
expect the systematic component of the differences to be correlated
between neighbouring histogram bins, we additionally consider the
cumulative $\chi^2/$d.o.f.\ difference for each histogram. That is, we
consider the $\chi^2/$d.o.f.\ obtained by adding to each histogram bin
the sum of all previous bins, adding the errors in quadrature. We
stress that these $\chi^2$ values are meant to be interpreted as a
measure of change relative to the statistical fluctuations in the
original predictions. They do not lend themselves to the more common
interpretations as goodness-of-fit measures or as $\chi^2$ hypothesis
tests.

The results are shown in tables~\ref{tab:chi2} and
\ref{tab:cum_chi2}. The largest $\chi^2$ values by far are found for
the W transverse momentum distributions at small transverse momenta,
see figure~\ref{fig:ptW} (left). While the absolute metric including
the W boson performs better than the absolute metric without W boson
in most bins, it shows the largest deviation in the first bin, leading
to the overall largest $\chi^2$ values. The deviations at small W
transverse momenta can be large because the bin width is small
compared to the maximum cell size. The relative metric with
reconstructed W boson performs best by a substantial margin. As
expected, the absolute metric with W boson stays closest to the
original prediction for large transverse momenta. The relative
distance measure again leads to the smallest overall deviations in the
$H_T$ distribution (figure \ref{fig:HT}), preserving the peak at small
values of $H_T$ better than the other metrics.

\begin{table}[htb]
  \centering
  \begin{tabular}{llll}
    \toprule
    & absolute & absolute + resonant & relative + resonant\\
    \midrule
    $\frac{d\sigma}{dp_{W\perp}}$ (figure \ref{fig:ptW}), $p_{W\perp} < 300\,$GeV& 1.87 & 4.30 & 0.0597\\
    $\frac{d\sigma}{dp_{W\perp}}$ (figure \ref{fig:ptW}), $p_{W\perp} > 300\,$GeV& 0.0248& 0.000202&0.00156\\
    $\frac{d\sigma}{dH_T}$ (figure~\ref{fig:HT})&1.06&0.801&0.489\\
    \bottomrule
\end{tabular}
\caption{$\chi^2/$d.o.f.\ for the various distributions and cell resampling metrics.}
  \label{tab:chi2}
\end{table}

\begin{table}[htb]
  \centering
  \begin{tabular}{llll}
    \toprule
    & absolute & absolute + resonant & relative + resonant\\
    \midrule
    $\frac{d\sigma}{dp_{W\perp}}$ (figure \ref{fig:ptW}), $p_{W\perp} < 300\,$GeV&1.76&4.09&0.0118\\
    $\frac{d\sigma}{dp_{W\perp}}$ (figure \ref{fig:ptW}), $p_{W\perp} > 300\,$GeV&0.0468&0.000555&0.00707\\
    $\frac{d\sigma}{dH_T}$ (figure~\ref{fig:HT})&1.45&1.38&0.638\\
    \bottomrule
\end{tabular}
\caption{Cumulative $\chi^2/$d.o.f.\ for the various distributions and cell resampling metrics.}
  \label{tab:cum_chi2}
\end{table}

To assess the improvement we first consider the negative-weight fraction $r_-$ and the Kish effective sample size fraction $f_{\text{ESS}}$~\cite{https://doi.org/10.1002/bimj.19680100122} proposed in~\cite{Farkh:2026mtw}:
\begin{equation}
  \label{eq:r-}
r_- = - \frac{\sum_{w_i < 0} w_i}{\sum_{i=1}^N \lvert w_i \rvert}, \qquad f_{\text{ESS}} = \frac{1}{N} \frac{\left(\sum_{i=1}^N w_i\right)^2}{\sum_{i=1}^N w_i^2},
\end{equation}
where $w_i$ is the weight of the $i$th event and $N$ the number of
events. As shown in table~\ref{tab:r-}, the conventional absolute
metric achieves the largest improvement, at the cost of larger changes
in the description of the W boson properties. After including the
reconstructed resonance in the metric, we still find a substantial
negative-weight reduction, leading to an increase of about three
orders of magnitude in the effective sample size.

\begin{table}[htb]
  \centering
  \begin{tabular}{lll}
    \toprule
    Sample & $r_-$ & $f_{\text{ESS}}$ \\
    \midrule
    original &    0.494 & $1.17 \times 10^{-9}$\\
    \texttt{cres} absolute &    0.352 & $2.29 \times 10^{-4}$\\
    \texttt{cres} absolute + resonant &    0.414 & $2.32 \times 10^{-6}$\\
    \texttt{cres} relative + resonant &    0.416 & $2.83 \times 10^{-6}$\\
    \bottomrule
  \end{tabular}
  \caption{Fractional contributions $r_-$ of negative weights to the total cross section and effective sample sizes $f_{\text{ESS}}$.}
  \label{tab:r-}
\end{table}

\begin{figure}[htb]
  \centering
  \includegraphics[width=0.45\linewidth]{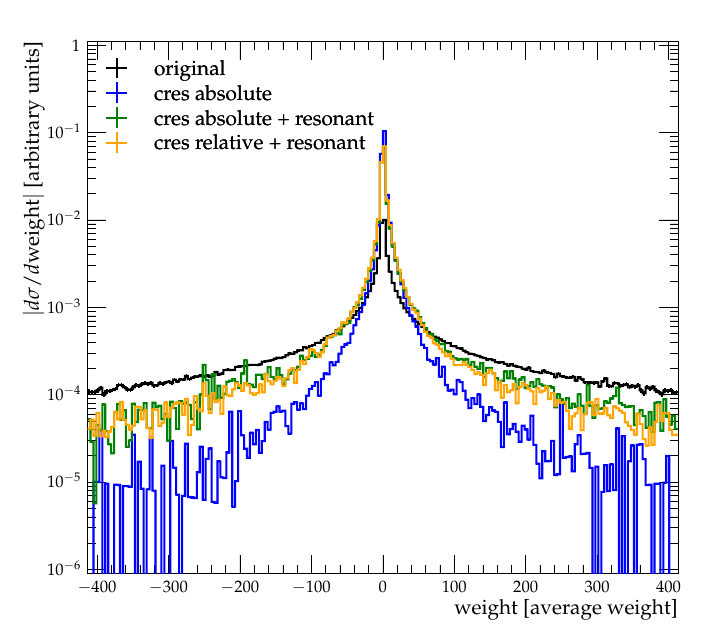}
  \caption{Event weight distribution before and after cell
    resampling. The original distribution (black) is compared to cell
    resampling using the conventional absolute metric (blue), a
    ``resonant'' metric including a reconstructed W boson (green), and
    the ``resonant + relative'' metric from section
    \ref{sec:metric_rel} (orange). Weights are normalised to the
    average weight in the original sample.}
  \label{fig:weights}
\end{figure}

In figure~\ref{fig:weights} we show the differential contribution from
negative and positive event weights to the cross section. A narrower
weight distribution is highly desirable, corresponding to faster
statistical convergence. All of the considered cell resampling
settings suppress the contribution from large absolute weights by at
least one order of magnitude. As expected, excluding the W boson from
the metric gives additional opportunities for weight cancellations, leading
to an even stronger negative-weight suppression for the original
absolute metric

\subsection{High-precision cell resampling}
\label{sec:precision_cres}

We now demonstrate that cell resampling with the newly proposed metric
can preserve predictions with very high precision while significantly
reducing the contribution from negative weights. To this end, we
restrict ourselves to the region of small W transverse momenta and
generate a high-statistics sample of $3.15\times 10^9$ events with
$p_{W\perp} < 55$\,GeV. All other settings as well as the analyses are
as in section~\ref{sec:metric_cmp}, c.f.\ also
table~\ref{tab:sherpa_param}.

In contrast to section~\ref{sec:metric_cmp}, we do not generate
Born-level and virtual-correction events separately, but instead
generate events with Born kinematics weighted with the combined Born
and virtual correction matrix elements. This enables partial
negative-weight cancellations already during event generation and
allows us to reach per cent-level accuracy or better in most bins with
the given sample size. For the cell resampling, we use the relative
metric including the W resonance defined in section
\ref{sec:metric_rel} with a maximum cell radius of $d(e, e') < 0.07$.

\begin{figure}[htb]
  \centering
  \includegraphics[width=0.45\linewidth]{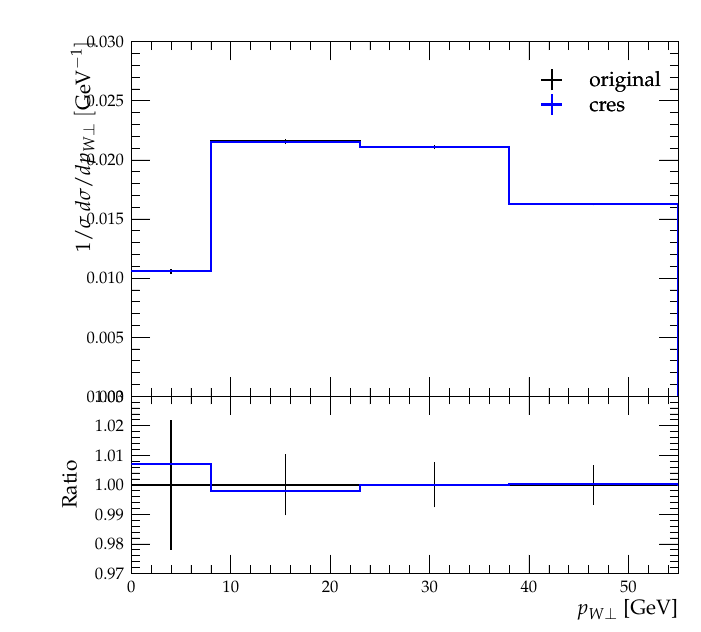}
  \includegraphics[width=0.45\linewidth]{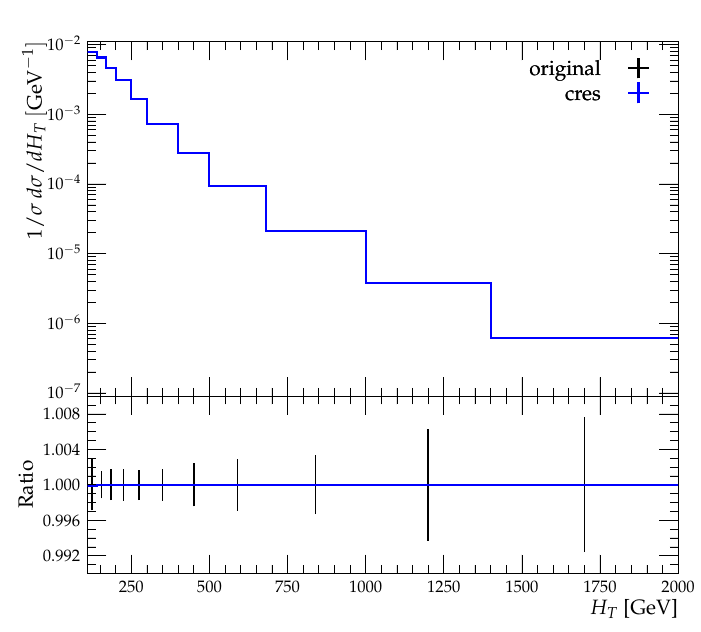}
  \caption{Predicted transverse momentum and $H_T$ distributions of the W boson
with $p_{W\perp} < 55$ GeV. The original distribution (black) is
compared to cell resampling using the ``resonant + relative'' metric
from section~\ref{sec:metric_rel} (blue).}
  \label{fig:ptW_HT_lowpt}
\end{figure}

As shown in figure~\ref{fig:ptW_HT_lowpt}, the W transverse momentum
distribution is preserved at the sub-per cent level, with the changes
due to cell resampling amounting to at most a third of the statistical
uncertainty. In the $H_T$ distribution we observe that the peak at
small values is no longer present when excluding events with large W
transverse momentum. Cell resampling preserves this distribution with
very high accuracy, with changes that are negligible compared to
the statistical uncertainty of a few per mille.

\begin{figure}[htb]
  \centering
  \includegraphics[width=0.45\linewidth]{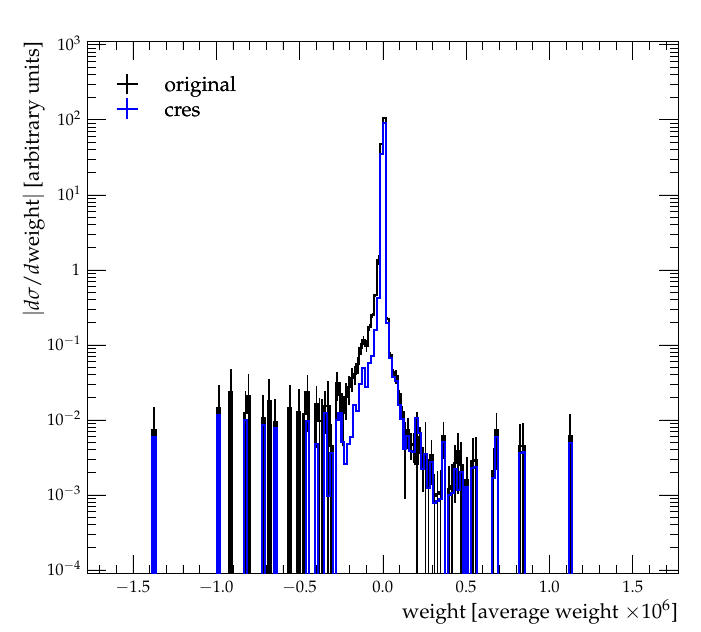}
  \caption{Event weight distribution before and after cell resampling
of a high-statistics sample with $p_{W\perp} < 55$ GeV. The original
distribution (black) is compared to cell resampling using the ``resonant
+ relative'' metric from section \ref{sec:metric_rel} (blue). Weights
are normalised to the average weight in the original sample.}
  \label{fig:weights_lowpt}
\end{figure}

In table~\ref{tab:r-_lowpt} and figure~\ref{fig:weights_lowpt} we
illustrate the cancellation of negative event weights, considering
again the negative weight fraction, the Kish effective sample size
fraction, and the weight distribution. We conclude that there is a
significant improvement, especially for large absolute weight values.

\begin{table}[htb]
  \centering
  \begin{tabular}{lcc}
    \toprule
    Sample & $r_-$ & $f_{\text{ESS}}$ \\
    \midrule
    original & 0.321 & $4.50\times 10^{-3}$ \\
    \texttt{cres} &    0.280 & $5.41\times 10^{-3}$ \\
    \bottomrule
  \end{tabular}
  \caption{Fractional contributions $r_-$ of negative weights to the
    total cross section and Kish effective sample size fraction
    $f_{\text{ESS}}$.}
  \label{tab:r-_lowpt}
\end{table}

\section{Conclusions}
\label{sec:conclusions}

In further developing the cell resampling approach to negative-weight
reduction we have presented a new phase-space metric. By including
reconstructed intermediate resonances in the distance computations the
sensitivity to their properties is significantly enhanced. To better
reflect experimental energy resolution uncertainties, we further
suggest to base the distance on relative differences in transverse
momentum. We aim for increased sensitivity in high-statistics regions
while enabling a better supression of negative weights in
low-statistics tails.

To assess the impact, we apply cell resampling with this new metric to
an event sample based on a NLO prediction for the production of a
leptonically decaying W boson together with two jets. Comparing to the
metric originally suggested in~\cite{Andersen:2021mvw}, we find that
the improved metric performs much better in the high-statistics region
of low transverse momenta. Even for large maximum cell sizes we find
small changes of about 1\%, well within the statistical
uncertainty. This demonstrates a marked improvement over the original
metric, which lead to much larger changes in this region. In all
cases the distribution of event weights is improved significantly,
with large absolute weights being suppressed by at least one order of
magnitude.

We further find that by choosing smaller maximum cell sizes
predictions can be preserved with extremely high accuracy, at the
sub-per cent or even sub-per mille level, while still achieving a
significant reduction in negative weights.

\acknowledgments

We thank Stephen Jones for collaboration in the early stages of this
work. The work of Jeppe R.~Andersen is supported by the STFC under
grant ST/X003167/1. Andreas Maier acknowledges financial support from
the Spanish Ministry of Science and Innovation (MICINN) through the
Spanish State Research Agency, under Severo Ochoa Centres of
Excellence Programme 2025-2029 (CEX2024001442-S). This work was
supported in part by the Spanish Ministry of Science and Innovation
(PID2020-112965GB-I00,PID2023-146142NB-I00), and by the Departament de
Recerca i Universities from Generalitat de Catalunya to the Grup de
Recerca 00649 (Codi: 2021 SGR 00649). This project has received
funding from the European Union’s Horizon 2020 research and innovation
programme under grant agreement No 824093. IFAE is partially funded by
the CERCA program of the Generalitat de Catalunya. This work was
performed in part at Aspen Center for Physics, which is supported by
National Science Foundation grant PHY-2210452.

\appendix

\section{Metric Properties of the Relative Distance}
\label{sec:metricproperties}

It is straightforward to observe that the relative distance $d_{\text{rel}}(p,p')$
defined in equation~\eqref{eq:dist_rel} fulfills $d_{\text{rel}}(p, p) = 0$ and
$d_{\text{rel}}(p, p') = d_{\text{rel}}(p', p)$. The condition $p \neq p': d_{\text{rel}}(p,p') > 0$
is also fulfilled, since each term is non-negative and $\beta \Delta R(p, p') = 0$  then implies
$ p_\perp \neq  p'_\perp$ and therefore $d_{\text{rel}}(p,p') = \alpha \left\lvert \log{\frac{p_\perp}{p'_\perp}}\right\rvert > 0$. Finally, to prove the triangle inequality $d_{\text{rel}}(p,p') \le d_{\text{rel}}(p,k) + d_{\text{rel}}(k,p')$ we assume without loss of generality that $p_\perp \geq p'_\perp$, such that
\begin{equation}
  \left\lvert \log\frac{p_\perp}{p'_\perp} \right\rvert
  = \log\frac{p_\perp}{p'_\perp}
    =\log\frac{p_\perp}{k_\perp} + \log\frac{k_\perp}{p'_\perp}
    \leq \left\lvert \log\frac{p_\perp}{k_\perp} \right\rvert + \left\lvert \log\frac{k_\perp}{p'_\perp}\right\rvert,
\end{equation}
which together with the well-known triangle inequality for $\Delta R$ completes the proof.

\bibliographystyle{JHEP}
\bibliography{papers}

\end{document}